\documentclass[a4paper,fleqn,usenatbib]{mnras}

\usepackage[T1]{fontenc}
\usepackage{ae,aecompl}

\usepackage{graphicx}	
\usepackage{color}
\usepackage{amsmath}	
\usepackage{amssymb}	

\title[Chemical abundance analysis of symbiotic giants]{Chemical abundance analysis of 13 southern symbiotic giants from
          high-resolution spectra at $\sim$1.56\,$\mu$m}

\author[C. Galan et al.]{
Cezary Ga{\l}an,$^{1}$\thanks{E-mail: cgalan@camk.edu.pl (CG)}
Joanna Miko{\l}ajewska,$^{1}$
Kenneth H. Hinkle,$^{2}$
Richard R. Joyce$^{2}$
\\
$^{1}$Nicolaus Copernicus Astronomical Center, Polish Academy of Sciences, Bartycka 18, PL-00-716 Warsaw, Poland\\
$^{2}$National Optical Astronomy Observatory, PO Box 26732, Tucson, AZ 85726, USA\\
}

\date{Accepted 2016 December 13. Received 2016 December 12; in original form 2016 September 8}

\pubyear{2016}

\begin{document}
\label{firstpage}
\pagerange{\pageref{firstpage}--\pageref{lastpage}}
\maketitle

\begin{abstract}
Symbiotic stars (SySt) are binaries composed of a\,star in the later stages
of evolution and a\,stellar remnant.  The enhanced mass-loss from the giant
drives interacting mass exchange and makes these systems laboratories for
understanding binary evolution.  Studies of the chemical compositions are
particularly useful since this parameter has strong impact on the
evolutionary path.
The previous paper in this series presented photospheric abundances for 24
giants in S-type SySt enabling a\,first statistical analysis.  Here, we
present results for an additional sample of 13 giants.  The aims are to
improve statistics of chemical composition involved in the evolution of
SySt, to study evolutionary status, mass transfer and to interpret this in
terms of Galactic populations.
High resolution, near-IR spectra are used, employing the spectrum synthesis
method in a\,classical approach, to obtain abundances of CNO and elements
around the iron peak (Fe, Ti, Ni).  Low resolution spectra in the region
around the \mbox{Ca\,{\sc ii}} triplet were used for spectral
classification.
The metallicities obtained cover a\,wide range with a\,maximum around
$\sim-0.2$\,dex.  The enrichment in the $^{14}$N isotope indicates that
these giants have experienced the first dredge-up.  Relative O and Fe
abundances indicate that most SySt belong to the Galactic disc; however, in
a\,few cases, the extended thick-disc/halo is suggested.  Difficult to
explain, relatively high Ti abundances can indicate that adopted
microturbulent velocities were too small by $\sim$0.2--0.3\,km\,s$^{-1}$ . 
The revised spectral types for V2905\,Sgr, and WRAY\,17-89 are M3 and M6.5,
respectively.
\end{abstract}




\section{Introduction}

Symbiotic stars (SySt) are among the longest orbital period interacting
binaries.  They are composed of a cool giant as the primary component and a
secondary that is a remnant of the latest stages of stellar evolution --
typically a white dwarf (WD) but a neutron star is also suggested in a few
cases.  The giant member is losing matter at a high rate, up to about
10$^{-7}$M\sun\,yr$^{-1}$, a rate systematically higher than for single
field giants \citep[][and references therein]{MIO2003}.  Interactions of the
mass flow with hard radiation from the hot, luminous remnant results in a
complex environment around each component of the system with many systems
being embedded in an ionized nebula.  The bulk of the giant's mass outflow
ultimately can be lost from the system.  However, a substantial part can be
accreted on to the compact object directly from the giant wind and/or via
Roche lobe overflow \citep{PoMa2007}.  When the system formed the current
remnant was the more massive component.  As this star evolved, mass was
transferred on to the star that is now the giant and traces of this mass
transfer may be detectable in the observed chemical composition of the
current red giant.  There is clear evidence for this history of binary
evolution in the S-type symbiotic systems.  Most systems have orbital
periods below $\sim$1000\,d and circularized orbits that indicate that mass
transfer and strong interactions took place before the present WD was formed
\citep{Mik2012}.

Chemical composition is secondary only to initial stellar mass in
determining stellar evolution.  In binary systems where abundances are
modified due to mass transfer, the information about single and binary
evolution becomes entangled.  However, abundances of some specific kinds of
chemical elements, eg.\,s-process elements produced during the AGB phase
and/or CNO abundances, combined with theoretical limitations and known basic
parameters of the stars in the system enable us to deduce the interactions  
and history of the mass exchange between components.  This motivated us to
determine chemical abundances in S-type symbiotic giants with the goal of
understanding their evolution.  Including new results presented here  
abundances are known for more than 40 red giants in SySt.  Statistical
analysis \citep[][Paper\,III]{Gal2016} addresses metallicity (from its proxy
the iron abundance), evolutionary status (by analysis of carbon and nitrogen
abundances and $^{12}$C/$^{13}$C ratio), and Galactic population membership
(by analysis of the alpha-element abundances, titanium and oxygen, relative
to iron).

In this paper, we present chemical compositions (abundances of CNO and
elements around the iron peak: Fe, Ti, Ni) for a sample of 13 red giants in
S-type symbiotic systems located in the Southern hemisphere.  The abundances
are measured using high-resolution near-IR spectra.  This is another step in
building a statistical base for future analysis of SySt samples representing
all stellar populations and various locations in the Milky Way and the
Magellanic Clouds.  The structure of the paper is as follows. Section\,2
lists and describes our spectroscopic observations. Section\,3 specifies the
methods used and presents the results.  The results are discussed briefly   
with comparison to our previous sample in Section\,4.


\begin{table}
  \caption{Journal of spectroscopic observations.}
\label{T1}
  \begin{tabular}{@{}l@{\hskip 3mm}c@{\hskip 3mm}c@{\hskip 3mm}c@{\hskip 3mm}c@{\hskip 3mm}c@{}}
  \hline
                      &  Id.$^{a}$ & Date      & HJD (mid)  & Phase$^{b}$& Range\\
                      &            & dd.m.yy   & $-2450000$ &            & [$\mu$m]\\
 \hline
 WRAY\,15-1470        &  63        & 24.5.10 & 5340.5687 & 0.23       & $\sim$1.56\\
 \hline
 Hen3\,1341           &  74        & 24.5.10 & 5340.6483 & 0.38       & $\sim$1.56\\
 \hline
 WRAY\,17-89          &  85        & 24.5.10 & 5340.6666 & --         & $\sim$1.56\\
                      &            & 23.3.16 & 7470.5698 & --         & $\sim$0.86\\
 \hline
 V2416\,Sgr           &  113       & 03.6.10 & 5350.6669 & --         & $\sim$1.56\\
 \hline
 V615\,Sgr            &  123       & 24.5.10 & 5340.9227 & 0.83       & $\sim$1.56\\
 \hline
 AS\,281              &  127       & 27.6.10 & 5374.7624 & 0.92       & $\sim$1.56\\
 \hline
 V2756\,Sgr           &  133       & 02.6.10 & 5349.8471 & 0.20       & $\sim$1.56\\
 \hline
 V2905\,Sgr           &  139       & 04.6.10 & 5351.9279 & 0.33       & $\sim$1.56\\
                      &            & 21.3.16 & 7468.6330 & 0.49       & $\sim$0.86\\
 \hline
 AR\,Pav              &  142       & 06.6.09 & 4988.9345 & 0.33       & $\sim$1.56\\
                      &            & 26.5.10 & 5342.7892 & 0.92       & $\sim$1.56\\
 \hline
 V3804\,Sgr           &  144       & 03.8.09 & 5046.7705 & 0.47       & $\sim$1.56\\
 \hline
 V4018\,Sgr           &  147       & 09.6.10 & 5356.6536 & 0.29       & $\sim$1.56\\
 \hline
 V919\,Sgr            &  159       & 03.8.09 & 5046.7861 & --         & $\sim$1.56\\
 \hline
 CD$-$43\degr14304    &  182       & 06.6.09 & 4988.8972 & 0.51       & $\sim$1.56\\
                      &            & 26.5.10 & 5342.7708 & 0.75       & $\sim$1.56\\
 \hline
 HR\,6020             &  --        & 21.3.16 & 7468.5838 & --         & $\sim$0.86\\
 \hline
 HD\,70421            &  --        & 22.3.16 & 7470.2712 & --         & $\sim$0.86\\
 \hline
 HR\,1247             &  --        & 22.3.16 & 7470.2880 & --         & $\sim$0.86\\
 \hline
 HR\,1264             &  --        & 22.3.16 & 7470.2954 & --         & $\sim$0.86\\
 \hline
 HR\,3718             &  --        & 22.3.16 & 7470.3270 & --         & $\sim$0.86\\
 \hline
 HR\,4902             &  --        & 22.3.16 & 7470.4247 & --         & $\sim$0.86\\
 \hline
 SW\,Vir              &  --        & 22.3.16 & 7470.4361 & --         & $\sim$0.86\\
 \hline
\end{tabular}   
\begin{list}{}{}
{\small{
\item[$^{a}$] Identification number according to \citet{Bel2000}.
\item[$^{b}$] Orbital phases that correspond to photometric minima (giant in
the front) are taken in majority from \citet{Gro2013}: WRAY\,15-1470
2451845+561$\times$E; Hen3\,1341 2451970+626$\times$E; V615\,Sgr
2452168+657$\times$E; AS\,281 2449021+533$\times$E; V2756\,Sgr
2451894+480$\times$E; V2905\,Sgr 2451630+508$\times$E; V3804\,Sgr
2451439+426$\times$E; V4018\,Sgr 2452129+513$\times$E;
from \citet{Sch2001}: 2448139+604.5$\times$E for AR\,Pav;
and for CD$-$43\degr14304 from ephemeris 2447015+1448$\times$E
recalculated according to circular orbit case \citep{Schm1998}.
}}
\end{list} 
\end{table}

\section{Observations and data reduction}

Near-IR spectra of 13 symbiotic SySt were observed in 2009 June, 2009
August, 2010 May and 2010 June using the Phoenix cryogenic echelle
spectrograph on the 8-m Gemini-S telescope.  The spectra were obtained
through a poor weather observing programme.  All are characterized by high
resolving power ($R\sim50000$) and most by high-S/N ratio ($\sim$\,100). 
The Gaussian instrumental profile is $\sim$6\,km\,s$^{-1}$ full width at
half-maximum (FWHM), corresponding to $\sim$0.31\AA\ at the observed
wavelength.  The spectra cover narrow spectral intervals ($\sim$60\AA)
located in wavelength range between 1.560--1.568\,{\rm \micron}.  This
region, free of telluric features, is dominated by first overtone OH lines
and numerous neutral atomic lines \mbox{Fe\,{\sc i}}, \mbox{Ti\,{\sc i}},
\mbox{Ni\,{\sc i}} plus a number of generally weak CN red system, $\Delta v
= -1$ lines and CO second-overtone vibration--rotation lines.  The selected
absorption lines are useful to measure abundances of carbon, nitrogen,
oxygen and elements around the iron peak: Ti, Fe and Ni.

\begin{figure}
   \includegraphics[width=\hsize]{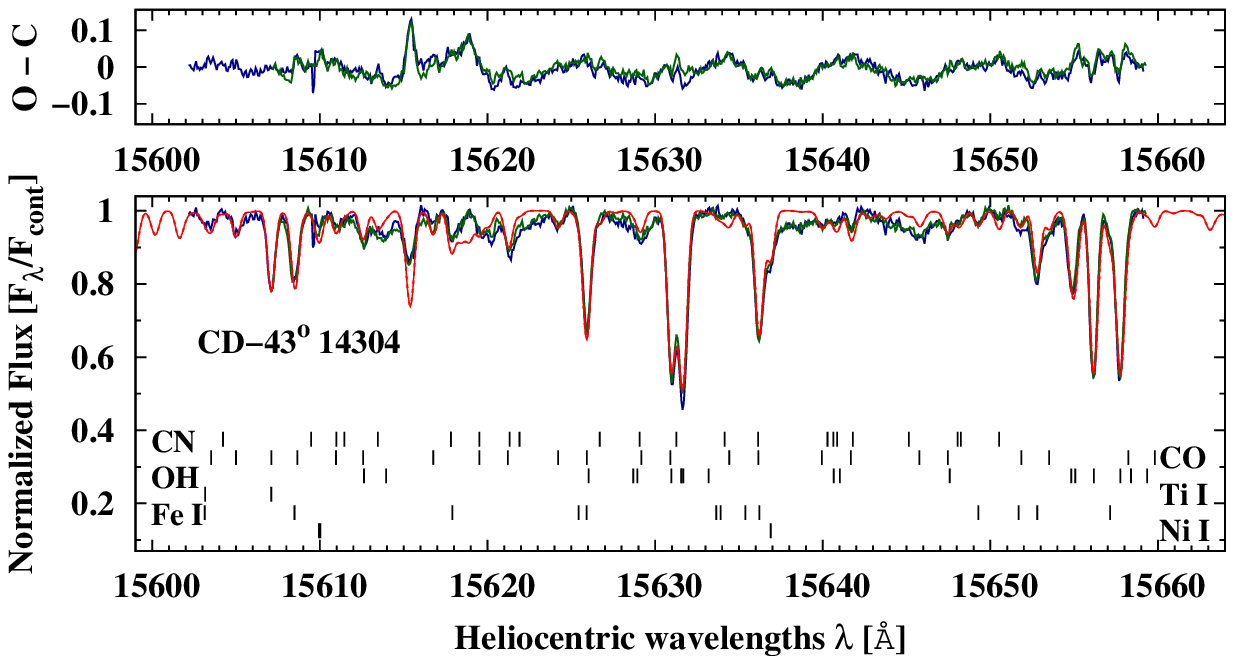}
      \caption{Spectra of CD$-$43\degr14304 observed 2009 June (blue line),
              2010 May (green line) and synthetic spectra (red continuous
              and dashed lines) calculated using the final abundances
              (Table\,\ref{T4}).
}
         \label{F1}
   \end{figure}

\begin{figure}
   \includegraphics[width=\hsize]{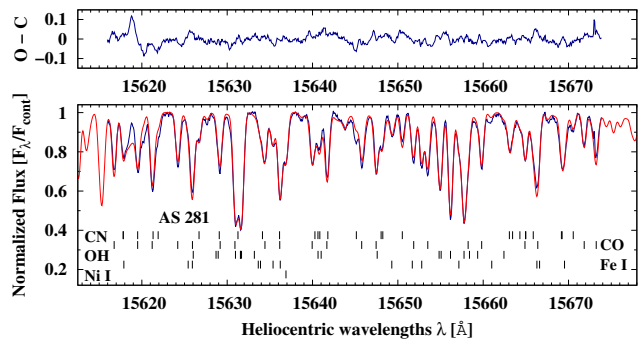}
      \caption{The spectrum of AS\,281 observed 2010 June (blue line) and a
              fitted synthetic spectrum (red line).}
         \label{F2}
   \end{figure}

The spectra were extracted from the raw data and wavelength calibrated using
standard reduction techniques \citep{Joy1992}.  The wavelength scales were
heliocentric corrected.  The {\small{IRAF}} cross-correlation program
{\small{FXCOR}} was employed \citep{Fit1993} to shift the observed spectra,
removing the stellar radial velocity, to the wavelength of the spectral
lines in synthetic spectra.  Representative spectra with synthetic fits are
shown in Figs\,\ref{F1}\,and\,\ref{F2}.  The spectra selected,
CD$-$43\degr14304 and AS\,281, are the most blueshifted and redshifted,
respectively, to show the whole explored wavelength range.

Lower resolution, very near-infrared spectra were also observed around the
\mbox{Ca\,{\sc ii}} triplet ($\lambda \sim$ 7200--9600 \AA\/, $R\sim2200$,
S/N $\sim 50$) for two symbiotic systems (V2905\,Sgr and WRAY\,17-89) and
several M-type giant standard stars with the Cassegrain spectrograph
(grating nr 11) operated on 1.9\,m 'Radcliffe' telescope in Sutherland
(South African Astronomical Observatory).  The observation took place in
2016 March as part of our project (JM, CG) to look for the evidences of
enhancement of giants with s-process elements from previous accretion. 
These spectra were used here to improve in classification of the spectral
type.  The journal of our spectroscopic observations is shown in
Table\,\ref{T1}.

%

%

%

\begin{table*}
\caption[]{Stellar parameters: effective temperature
$T_{\rm{eff}}$ and surface gravity $\log{g}$.}
\label{T2}
{\small{
\begin{tabular}{@{}l@{\hskip 3mm}l@{\hskip 3mm}l@{\hskip 3mm}l@{\hskip 3mm}l@{\hskip 3mm}l@{\hskip 3mm}l@{\hskip 3mm}l@{\hskip 3mm}l@{\hskip 3mm}l@{\hskip 3mm}l@{\hskip 3mm}l@{}}
\hline
                    & Spec.        & $T_{\rm{eff}}^{[1]}$ & $T_{\rm{eff}}^{[2]}$ & $J-K^{[3,4]}$ & $E$($B-V$)$^{[5]}$   & ($J-K$)$_0$      & $T_{\rm{eff}}^{[6]}$ & $\log{g}$$^{[6]}$ & $\log{g}$$^{[7]}$ & $T_{\rm{eff}}^{a}$ & $\log{g}^{a}$ \\
                    & Type         & [K]                  & [K]                  & mag           & mag                  & mag              & [K]                  &                   &                   & [K]                &               \\
  \hline
WRAY\,15-1470       & M3$^{[8]}$   & 3560$\pm$75          & 3586                 & 1.38$\pm$0.07 & $<$0.49$\pm$0.02     & $>$1.15$\pm$0.07 & $<$3570$\pm$140      & $<$+0.5$\pm$0.3   & 0.5--0.8          & 3600               & 0.5           \\
Hen\,3-1341         & M4$^{[8]}$   & 3460$\pm$75          & 3476                 & 1.32$\pm$0.07 & $<$0.31$\pm$0.01     & $>$1.17$\pm$0.09 & $<$3520$\pm$180      & $<$+0.4$\pm$0.3   & 0.4--0.7          & 3500               & 0.5           \\
WRAY\,17-89         & M6.5$^{[11]}$& 3170$\pm$80          & 3203                 & 2.06$\pm$0.07 & $<$1.49$\pm$0.05     & $>$1.33$\pm$0.13 & $<$3200$\pm$260      & $<$-0.1$\pm$0.5   & 0.0--0.3          & 3200               & 0.0           \\
V2416\,Sgr          & M6$^{[8]}$   & 3240$\pm$75          & 3258                 & 1.99$\pm$0.08 & --                   & --               & --                   & --                & 0.1--0.4          & 3300               & 0.0           \\
V615\,Sgr           & M5.5$^{[8]}$ & 3300$\pm$75          & 3312                 & 1.32$\pm$0.06 & $<$0.15$\pm$0.01     & $>$1.25$\pm$0.07 & $<$3360$\pm$150      & $<$+0.1$\pm$0.3   & 0.2--0.5          & 3300               & 0.0           \\
AS\,281             & M6$^{b}$     & 3240$\pm$75          & 3258                 & 1.49$\pm$0.06 & $<$0.48$\pm$0.03     & $>$1.26$\pm$0.10 & $<$3340$\pm$200      & $<$+0.1$\pm$0.4   & 0.1--0.4          & 3300               & 0.0           \\
V2756\,Sgr          & M4$^{b}$     & 3460$\pm$75          & 3476                 & 1.34$\pm$0.06 & $<$0.27$\pm$0.01     & $>$1.21$\pm$0.08 & $<$3440$\pm$160      & $<$+0.3$\pm$0.3   & 0.4--0.7          & 3500               & 0.5           \\
V2905\,Sgr          & M3$^{[11]}$  & 3560$\pm$75          & 3586                 & 1.26$\pm$0.07 & $<$0.28$\pm$0.01     & $>$1.13$\pm$0.08 & $<$3610$\pm$170      & $<$+0.6$\pm$0.3   & 0.5--0.8          & 3600               & 0.5           \\
AR\,Pav             & M5$^{[8]}$   & 3355$\pm$75          & 3367                 & 1.17$\pm$0.07 & $<$0.08$\pm$0.01     & $>$1.13$\pm$0.08 & $<$3610$\pm$170      & $<$+0.6$\pm$0.3   & 0.3--0.6          & 3400               & 0.0           \\
V3804\,Sgr          & M5$^{[9]}$   & 3355$\pm$75          & 3367                 & 1.46$\pm$0.06 & $<$0.21$\pm$0.01     & $>$1.35$\pm$0.07 & $<$3150$\pm$130      & $<$-0.2$\pm$0.2   & 0.3--0.6          & 3300               & 0.0           \\
V4018\,Sgr          & M4$^{[10]}$  & 3460$\pm$75          & 3476                 & 1.24$\pm$0.06 & $<$0.34$\pm$0.01     & $>$1.09$\pm$0.08 & $<$3690$\pm$160      & $<$+0.7$\pm$0.3   & 0.4--0.7          & 3500               & 0.5           \\
V919\,Sgr           & M4.5$^{[8]}$ & 3410$\pm$75          & 3421                 & 1.38$\pm$0.07 & $<$0.21$\pm$0.01     & $>$1.27$\pm$0.07 & $<$3310$\pm$150      & $<$+0.1$\pm$0.3   & 0.3--0.6          & 3400               & 0.5           \\
CD$-$43\degr14304   & K7$^{[8]}$   & $\sim$3910           & $\sim$3980           & 1.08$\pm$0.09 & $<$0.03$\pm$0.01     & $>$1.07$\pm$0.09 & $<$3740$\pm$190      & $<$+0.8$\pm$0.4   & --                & 3900               & 1.0           \\
  \hline
\end{tabular}
}}
\begin{list}{}{}
\small{
\item[References:] total Galactic extinction according to
$^{[5]}$\citet{Sch2011} and \cite{Schl1998}, infrared colours from 2MASS
$^{[3]}$\citep{Phi2007} transformed to $^{[4]}$\citet{BeBr1988} photometric
system.  Spectral types are from: $^{[8]}$\citet{MuSm1999},
$^{[9]}$ \citet{MTS1995}, $^{[10]}$ \citet{All1980}, $^{[11]}$ this work
\item[Callibration:] $^{[1]}$\citet{Ric1999}, $^{[2]}$\citet{VBe1999},
$^{[6]}$\citet{Kuc2005}, $^{[7]}$\citet{DuSc1998}.
\item[$^{a}$]adopted.
\item[$^{b}$]spectral types changed to later by one spectral subclass relative to
\citet{MTS1995} spectral classification.
}
\end{list}
\end{table*}

\section{Analysis and results}\label{AR}

To measure chemical abundances we used spectral synthesis techniques
employing local thermodynamic equilibrium (LTE) analysis.  The model
atmospheres were from a 1D, hydrostatic {\small{MARCS}} model atmospheres
\citep{Gus2008}.  This analysis was also employed in previous papers
(\citealt[][(Paper\,I)]{Mik2014}, \citealt[][(Paper\,II)]{Gal2015},
\citealt[][(Paper\,III)]{Gal2016}).  Justifications for the adopted
methodology can be found there.  The excitation potentials and $gf$-values
for transitions were taken in the case of atomic lines from the list by
\citet{MeBa1999} and for the molecular data we used line lists by
\citet{Goo1994} for CO, by \citet{Kur1999} for OH and by \citet{Sne2014} for
CN.  The {\small{WIDMO}} code developed by M.\,R.\,Schmidt \citep{Sch2006}
was used to calculate synthetic spectra.

\begin{table}
  \caption{Spectral classification of V2905\,Srg, and WRAY\,17-89 based on
          the strength of five TiO band heads.}
\label{T3}
  \begin{tabular}{@{}l@{\hskip 3mm}l@{\hskip 3mm}l@{\hskip 3mm}l@{\hskip 3mm}l@{\hskip 3mm}l@{\hskip 3mm}l@{}}
  \hline
                & \multicolumn{5}{c}{TiO band head ($\lambda$ [\AA]).} &      \\
                & 7589 & 8194     & 8432       & 8859  & 9209          & Mean \\
  \hline
 V2905\,Sgr     & M3   & $\leq$M2 & M4--3.5    & $<$M3 & $\leq$M4      & M3   \\
 WRAY\,17-89    & M5   & M6       & M5.5$^{a}$ & M6    & M7            & M6   \\
  \hline
\end{tabular}   
\begin{list}{}{}
\small{
\item[$^a$] can be significantly shallowed by the emission in \mbox{Ca\,{\sc ii}} triplet.
}
\end{list}
\end{table}

In summary the abundances calculations were performed as follows.  The
starting values of abundances were obtained by fitting by eye, through
several iterations, alternately from molecular and atomic lines.  Next, the
simplex algorithm \citep{Bra1998} was used for $\chi^2$ minimization with
seven free parameters: C, N, O, Fe, Ti, Ni abundances and the rotational
velocity.  The rotational velocity must in this case be treated as a free
parameter because there was no way to measure reliably its value from
severely blended spectra.  A number of atmosphere models have been tested
with the above procedure for each target, when needed, to choose the one
with the best-matching metallicity.  The stellar parameters, effective
temperature ($T_{\rm{eff}}$) and surface gravity ($\log{g}$), could not be
measured directly from our spectra because of lack the lines from ionized
elements and unblended lines with different intensities for the same
elements.  The atmospheric parameters needed for selecting the most suitable
atmosphere model were instead derived from the spectral types
(Table\,\ref{T2}).  In most cases, the spectral types were derived from TiO
bands in the near-IR with high accuracy of approximately one spectral
subclass \citep{MuSm1999}.

   \begin{figure}
   \includegraphics[width=\hsize]{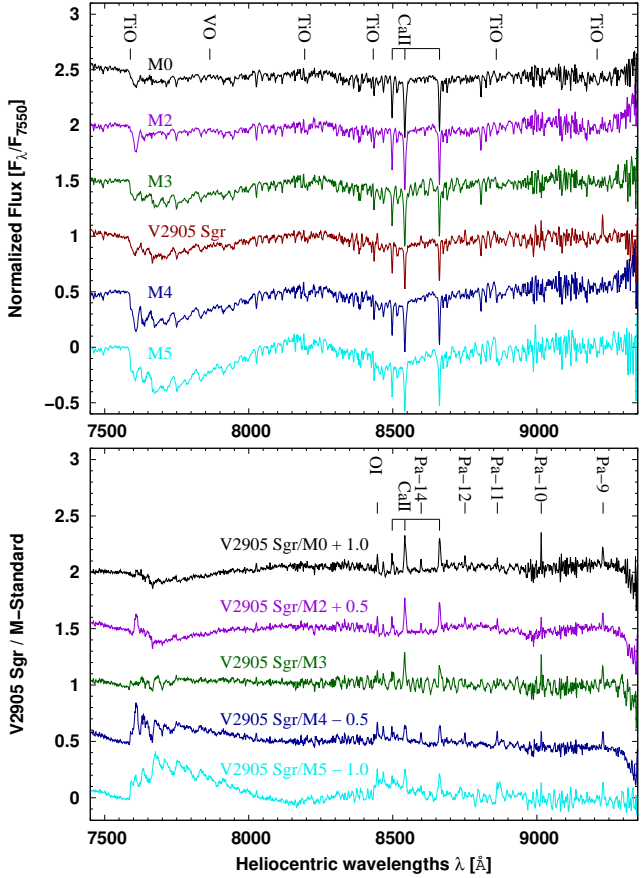}
      \caption{The spectrum of V2905\,Sgr (red) is compared with the spectra
of spectroscopic standards ({\sl top}) that are giants of types from M5 to
M0.  At the {\sl bottom} ratios of the V2905\,Sgr spectrum with the spectra
of standards.  The spectra are shifted by 0.5 for clarity.  The flattest
spectrum around the value one results from division by the M3 standard
spectrum (green).  The contribution from nebular continuum is seen in
emission from \mbox{Ca\,{\sc ii}} triplet, OI ($\lambda$ 8442.36\,\AA\/)
and numerous lines of the Paschen series.}
         \label{F3}
   \end{figure}

   \begin{figure}
   \includegraphics[width=\hsize]{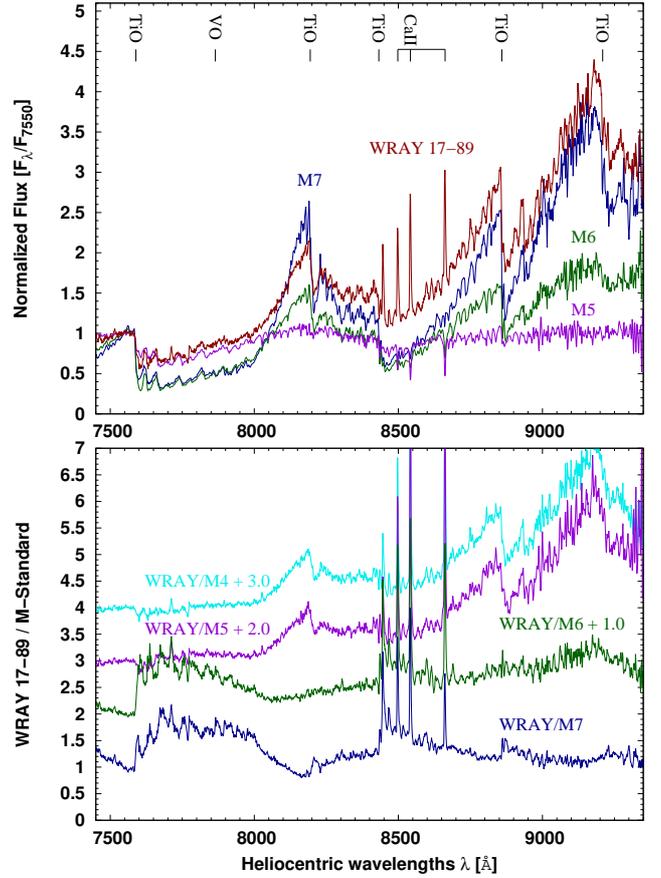}
      \caption{The spectrum of WRAY\,17-89 (red) is compared with the
spectra of spectroscopic standards ({\sl top}) that are giants of spectral
types M7, M6 and M5.  At the {\sl bottom} the ratios of the WRAY\,17-89
spectrum with the spectra of standards are presented (the spectra are
shifted by 1.0 for clarity).}
         \label{F4}
   \end{figure}

V2905\,Sgr and WRAY\,17-89 were classified by \citet{Mik1997} as giants of
spectral types M0 and M3, respectively.  We found that this classification
based on optical spectra resulted in a too early spectral type.  We used
near-infrared spectra around the \mbox{Ca\,{\sc ii}} triplet to improve the
spectral type classification.  This domain is particularly useful in
studying properties of the late-type stars in SySt because they are
intrinsically bright in this range and contamination by the nebula and hot
component is negligible.  To determine the spectral types, we used a method
similar to \citet{MuSm1999} with two approaches.  (i) In one approach, the
strength of TiO features in the spectrum of SySt were compared by eye with
those in the spectra of M-type giant standard stars with well-known spectral
types: HR\,3718 (M0, \citet{MuSm1999}), HR\,1247, HR\,4902, HR\,1264,
HR\,6020, SW\,Vir (M2, M3, M4, M5 and M7, respectively \citet{Fea1990}) and
HD\,70421 (M6, \citet{ScLa1988}).  Several TiO bands with the heads at
$\sim$ 7589, 8194, 8432, 8859 and 9209 \AA\/ are especially useful.  (ii) In
the second approach, the SySt spectrum divided by the standard spectrum is
assessed for the smoothest ratio.  Interestingly, the \mbox{Ca\,{\sc ii}}
triplet itself is a diagnostic of temperature and luminosity class as well
as metallicity for 'normal', singular red giants of early spectral types. 
However, it does not work for giants interacting in symbiotic systems.  In
the symbiotics the \mbox{Ca\,{\sc ii}} lines are frequently modified by
strong contribution from nebula, as clearly seen in the case of V2905\,Sgr
(Fig.\ref{F3}, {\sl bottom}).

For V2905\,Sco, the spectral types from the depths of TiO band heads are in
the range M2--M4 with the average around M3 (Table\,\ref{T3}).  The strong
TiO band with the head at $\lambda \sim 7589 \AA$ clearly indicates an M3
type (Fig.\,\ref{F3}, {\sl top}).  The mean spectra resulting from
division of V2905\,Sgr spectrum by the spectral standards achieve the
smallest deviations from flatness for the spectral type M3
(Fig.\,\ref{F3}, {\sl bottom}).  A visual comparison of the spectrum of
V2905\,Sgr to the standards shows that the spectrum is a best match to M3 at
shorter wavelengths and in the long wavelength range is earlier than M4. 
Therefore, we adopt M3 as the most suitable classification for this giant.

For WRAY\,17-89 the depths of TiO bands indicate an M6 spectral type
(Table\,\ref{T3}, and Fig.\,\ref{F4}, {\sl top}).  The result of division
of the WRAY\,17-89 spectrum by the spectral standards gives the smallest
deviation from flatness for M7 (Fig.\,\ref{F4}, {\sl bottom}).  From a
visual comparison, the spectrum at longer wavelengths resemble an M7 giant
while at the short wavelength M6-5 is a better match.  We adopted M6.5-type
for this star.

Adopting M3 and M6.5 spectral types for the giants in V2905\,Sgr, and
WRAY\,17-89, respectively, gives excellent agreement between the effective
temperatures obtained from the spectral types and those estimated from
infrared colours (Table\,\ref{T2}).

In the case of two other objects, AS\,281 and V2756\,Sgr, we found that the
spectral types by \citet{MTS1995}, M5 and M3 respectively, seem too early. 
We do not have 7200--9600 \AA\ spectra for these objects.  However, we
reanalyzed the published spectra looking at the strength of TiO bands heads
and the VO band head at 7865\,\AA.  We have reclassified the spectral types
by one subclass to M6 and M4, respectively.  The method of classification by
\citet{MTS1995} typically leads to earlier spectral types than the
classification by \citet{MuSm1999} for the same objects.  For this reason,
we previously \citep{Gal2016} reclassified the spectral type of SS73\,96.

To derive $T_{\rm{eff}}$, the calibrations of \citet{Ric1999} and
\citet{VBe1999} were used.  The surface gravity was estimated using two
methods.  In the first technique, upper limits to $\log{g}$ were found from
the $T_{\rm{eff}}$--$\log{g}$--colour relation for late-type giants
\citep{Kuc2005}.  Infrared intrinsic colours were derived from known $J$ and
$K$ magnitudes \citep{Phi2007} corrected for extinction by adopting colour
excesses from the \citet{Sch2011} maps of the Galactic extinction.  The
temperatures derived from spectral types are in general within the limits on
the temperatures derived from dereddened ($J-K$)$_0$ colours.  The second
method derives the radius of the giant star from the radius--spectral type
relation \citet[Table\,2]{DuSc1998} by assuming that the typical mass of
giant in the S-type symbiotic systems is $\sim$1--2 M\sun\, \citep{Mik2003}. 
Values of $T_{\rm{eff}}$ and $\log{g}$ for the model atmospheres used to
synthesize the spectra are shown in the rightmost columns of
Table\,\ref{T2}.

AR\,Pav is an eclipsing binary so an additional constraint can be placed on
the $\log{g}$ value.  With the orbital inclination known \citet{Qui2002}
were able to solve for the mass and radius from the solution to the red
giant's orbital parameters.  The giant's mass is
$M_{\rm{rg}}=2.5\pm0.6$M\sun.  The giant's radius is 137$\pm20$R\sun\, if
the orbit is seen edge on and ranges up to $\sim210$R\sun\, for the maximum
allowed inclination, $i \approx 70$\degr.  The giant nearly fills its
Roche lobe so is assumed to be tidally distorted.  The resulting range for
the surface gravity is $\log{g}\approx$0.1--0.8.  Observations of infrared
light variation resulting from ellipsoidal changes suggest that the giant is
tidally distorted with $R_{\rm{rg}}=\sim190$R\sun\, \citet{Rut2007}.  With
this value $\log{g}=0$, close to the lower limit.  This is in good agreement
with values for other symbiotic giants listed in Table\,\ref{T2}.

   \begin{figure}
   \includegraphics[width=\hsize]{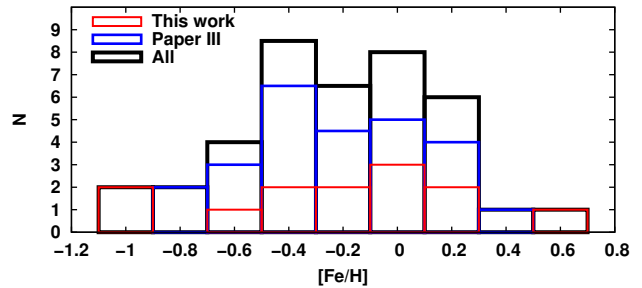}
      \caption{The distribution for the number 'N' of objects, counted at
0.2\,dex intervals, as a function of metallicity ([Fe/H]) for all symbiotic
giants studied by us so far including two samples (black): the current
sample (red) based on $H$-band spectra only, and previous sample (blue)
obtained from analysis of both $H$- and $K$-band regions (Paper\,III).}
         \label{F5}
   \end{figure}

The parameters describing the atmospheric motion, the micro ($\xi_{\rm t}$)
and macro ($\zeta_{\rm t}$) turbulence velocities, were set to typical
values for cool Galactic red giants.  The most commonly used values are
$\xi_{\rm{t}} = 2$ km\,s$^{\rm{-1}}$ and $\zeta_{\rm{t}}$ =
3\,km\,s$^{\rm{-1}}$ (see Paper\,III).

\begin{table*}
  \caption{Final abundances derived on the scale of $\log{\epsilon}(X) =
\log{(N(X) N(H)^{-1})} + 12.0$, abundances relative to solar abundances
together with the formal fitting errors\,$^a$, and rotational velocities.}
  \label{T4}
{\small{
  \begin{tabular}{@{}lrrrrrrrr@{}}
  \hline
  \hline
                    & C                   & N         & O              & Ti             & Fe             & Ni             & $V_{\rm{rot}} \sin{i}$   & $V_{\rm{rot}} \sin{i}$   \\
                    & $\log{\epsilon(X)}$ &           &                &                &                &                & $[\rm{km}\,\rm{s}^{-1}]$ & $[\rm{km}\,\rm{s}^{-1}]$ \\
                    & [$X$]$^b$           &           &                &                &                &                & (FIT)                    & (CCF)                    \\
  \hline
WRAY\,15-1470       & $ 8.03\pm0.02$ & $ 7.77\pm0.04$ & $ 8.77\pm0.01$ &   --           & $ 7.21\pm0.04$ & $ 5.83\pm0.07$ & 5.4$\pm$0.4              & 4.5$\pm$0.8              \\
                    & $-0.40\pm0.07$ & $-0.06\pm0.09$ & $+0.08\pm0.06$ &   --           & $-0.26\pm0.08$ & $-0.37\pm0.11$ & &\\
Hen\,3-1341         & $ 7.90\pm0.01$ & $ 7.75\pm0.02$ & $ 8.53\pm0.01$ & $ 4.80\pm0.06$ & $ 6.93\pm0.02$ & $ 5.75\pm0.04$ & 6.6$\pm$0.1              & 5.3$\pm$0.4              \\
                    & $-0.53\pm0.06$ & $-0.08\pm0.07$ & $-0.16\pm0.06$ & $-0.13\pm0.10$ & $-0.54\pm0.06$ & $-0.45\pm0.08$ & &\\
WRAY\,17-89         & $ 8.86\pm0.03$ & $ 9.05\pm0.11$ & $ 9.78\pm0.04$ & $ 5.56\pm0.18$ & $ 8.14\pm0.05$ & $ 6.15\pm0.12$ & 9.5$\pm$0.2              & 6.3$\pm$0.4              \\
                    & $+0.43\pm0.08$ & $+1.22\pm0.16$ & $+1.09\pm0.09$ & $+0.63\pm0.22$ & $+0.67\pm0.09$ & $-0.05\pm0.16$ & &\\
V2416\,Sgr          & $ 8.43\pm0.03$ & $ 8.57\pm0.10$ & $ 9.05\pm0.02$ & $ 5.54\pm0.09$ & $ 7.67\pm0.06$ & $ 6.29\pm0.09$ & 7.7$\pm$0.3              & 6.1$\pm$0.6              \\
                    & ~$0.00\pm0.08$ & $+0.74\pm0.15$ & $+0.36\pm0.07$ & $+0.61\pm0.13$ & $+0.20\pm0.10$ & $+0.09\pm0.13$ & &\\
V615\,Sgr           & $ 7.91\pm0.04$ & $ 8.30\pm0.12$ & $ 8.66\pm0.03$ & $ 5.06\pm0.12$ & $ 7.40\pm0.03$ & $ 6.16\pm0.11$ & 7.0$\pm$0.5              & 6.0$\pm$0.5              \\
                    & $-0.52\pm0.09$ & $+0.47\pm0.17$ & $-0.03\pm0.08$ & $+0.13\pm0.16$ & $-0.07\pm0.07$ & $-0.04\pm0.15$ & &\\
AS\,281             & $ 7.94\pm0.04$ & $ 7.53\pm0.10$ & $ 8.27\pm0.01$ &   --           & $ 7.11\pm0.04$ & $ 6.23\pm0.12$ & 6.1$\pm$0.3              & 5.7$\pm$0.8              \\
                    & $-0.49\pm0.09$ & $-0.30\pm0.15$ & $-0.42\pm0.06$ &   --           & $-0.36\pm0.08$ & $+0.03\pm0.16$ & &\\
V2756\,Sgr          & $ 8.23\pm0.02$ & $ 7.96\pm0.07$ & $ 8.81\pm0.04$ & $ 4.85\pm0.10$ & $ 7.39\pm0.03$ & $ 6.03\pm0.09$ & 8.0$\pm$0.5              & 6.9$\pm$0.6              \\
                    & $-0.20\pm0.07$ & $+0.13\pm0.12$ & $+0.12\pm0.09$ & $-0.08\pm0.14$ & $-0.08\pm0.07$ & $-0.17\pm0.13$ & &\\
V2905\,Sgr          & $ 7.33\pm0.04$ & $ 7.41\pm0.06$ & $ 7.89\pm0.04$ &   --           & $ 6.44\pm0.03$ & $ 5.50\pm0.09$ & 9.7$\pm$0.6              & --                       \\
                    & $-1.10\pm0.09$ & $-0.42\pm0.11$ & $-0.80\pm0.09$ &   --           & $-1.03\pm0.07$ & $-0.70\pm0.13$ & &\\
AR\,Pav             & $ 7.86\pm0.02$ & $ 7.91\pm0.05$ & $ 8.51\pm0.02$ & $ 4.51\pm0.18$ & $ 7.22\pm0.04$ & $ 5.82\pm0.09$ &11.1$\pm$0.7              &10.5$\pm$0.6               \\
                    & $-0.58\pm0.07$ & $+0.08\pm0.10$ & $-0.19\pm0.07$ & $-0.43\pm0.22$ & $-0.26\pm0.08$ & $-0.38\pm0.13$ & 8.7$\pm$0.4              & 8.1$\pm$1.4               \\
V3804\,Sgr          & $ 8.16\pm0.01$ & $ 7.89\pm0.03$ & $ 8.60\pm0.02$ & $ 4.81\pm0.06$ & $ 7.52\pm0.01$ & $ 5.84\pm0.04$ &10.0$\pm$0.2              & 8.2$\pm$0.3               \\
                    & $-0.27\pm0.06$ & $+0.06\pm0.08$ & $-0.09\pm0.07$ & $-0.12\pm0.10$ & $+0.05\pm0.05$ & $-0.36\pm0.08$ & &\\
V4018\,Sgr          & $ 8.48\pm0.02$ & $ 8.26\pm0.03$ & $ 9.00\pm0.02$ & $ 5.16\pm0.08$ & $ 7.67\pm0.02$ & $ 6.40\pm0.07$ & 8.5$\pm$0.3              & 7.5$\pm$0.9              \\
                    & $+0.05\pm0.07$ & $+0.43\pm0.08$ & $+0.31\pm0.07$ & $+0.23\pm0.12$ & $+0.20\pm0.06$ & $+0.20\pm0.11$ & &\\
V919\,Sgr           & $ 7.98\pm0.01$ & $ 7.73\pm0.03$ & $ 8.41\pm0.01$ & $ 4.76\pm0.06$ & $ 7.10\pm0.03$ & $ 5.90\pm0.03$ & 6.6$\pm$0.2              & 6.8$\pm$0.3              \\
                    & $-0.45\pm0.06$ & $-0.10\pm0.08$ & $-0.28\pm0.06$ & $-0.17\pm0.10$ & $-0.37\pm0.07$ & $-0.30\pm0.07$ & &\\
CD$-$43\degr14304   & $ 7.40\pm0.02$ & $ 7.33\pm0.05$ & $ 8.32\pm0.01$ & $ 4.37\pm0.04$ & $ 6.54\pm0.02$ & $ 5.41\pm0.03$ & 5.3$\pm$0.2              & 5.9$\pm$0.8              \\
                    & $-1.03\pm0.07$ & $-0.50\pm0.10$ & $-0.37\pm0.06$ & $-0.56\pm0.08$ & $-0.93\pm0.06$ & $-0.79\pm0.07$ & 5.2$\pm$0.2              & 5.1$\pm$1.3              \\
  \hline
Sun                 & $ 8.43\pm0.05$ & $ 7.83\pm0.05$ & $ 8.69\pm0.05$ & $ 4.93\pm0.04$ & $ 7.47\pm0.04$ & $ 6.20\pm0.04$ & &\\
  \hline
  \hline
\end{tabular}
}}
\begin{list}{}{}
\small{
\item[$^a$] 3$\sigma$.
\item[$^b$] Relative to the Sun [$X$] abundances in respect to the solar composition of \citet{Asp2009} and \citet{Sco2015}.
}
\end{list}
\end{table*}

\begin{table}
  \caption{Sensitivity of abundances to uncertainties in the stellar
parameters for cool M-type giants ($T_{\rm{eff}} \leq 3600$\,K) at the {\sl
Top}, and for yellow symbiotic CD$-$43\degr14304 ($T_{\rm{eff}} \sim
4000$\,K) at the {\sl Bottom}.}
\label{T5}  
  \begin{tabular}{@{}lrrrr@{}}

  \hline
  \hline
  \multicolumn{5}{c}{cool M-type giants ($T_{\rm{eff}} \leq 3600$\,K)}\\
  \hline
  $\Delta X$ & $\Delta T_{\rm{eff}} = +100$\,K & $\Delta \log{g} = +0.5$ & $\Delta  \xi_{\rm{t}} = +0.25$ & $\Delta^a$\\
  \hline
  C          & $+0.04$ & $+0.22$ & $-0.03$ & $\pm$0.22 \\
  N          & $+0.03$ & $+0.01$ & $-0.06$ & $\pm$0.06 \\
  O          & $+0.12$ & $+0.07$ & $-0.05$ & $\pm$0.15 \\
  Ti         & $+0.05$ & $+0.14$ & $-0.20$ & $\pm$0.25 \\
  Fe         & $-0.04$ & $+0.14$ & $-0.08$ & $\pm$0.16 \\
  Ni         & $-0.06$ & $+0.17$ & $-0.10$ & $\pm$0.20 \\
  \hline
  \hline
  \multicolumn{5}{c}{hot yellow giant CD$-$43\degr14304 ($T_{\rm{eff}} \sim$ 3900 -- 4250\,K)}\\
  \hline
  $\Delta X$ & $\Delta T_{\rm{eff}} = +100$\,K & $\Delta \log{g} = +0.5$ & $\Delta  \xi_{\rm{t}} = +0.25$ & $\Delta^a$\\
  \hline
   C         & $+0.05$ & $+0.20$ &~~$0.00$ & $\pm$0.20 \\
   N         & $+0.11$ & $-0.07$ & $-0.04$ & $\pm$0.13 \\
   O         & $+0.22$ & $-0.02$ & $-0.04$ & $\pm$0.22 \\
   Ti        & $+0.15$ & $+0.03$ & $-0.03$ & $\pm$0.16 \\
   Fe        & $+0.01$ & $+0.07$ & $-0.05$ & $\pm$0.08 \\
   Ni        & $+0.00$ & $+0.09$ & $-0.02$ & $\pm$0.09 \\
  \hline
\end{tabular}   
\begin{list}{}{}
\item[$^a$]$[(\Delta T_{\rm{eff}})^2 + (\Delta \log{g})^2 + (\Delta \xi_{\rm{t}})^2]^{0.5}$.
\end{list} 
\end{table}

The final abundances calculated for 13 S-type symbiotic systems are shown in
Table\,\ref{T4}.  The table also contains rotational velocities compared
with values obtained via the cross-correlation technique (CCF).  Fits to the
data with synthetic spectra for CD$-$43\degr14304 and AS\,281 are shown in
Figs\,\ref{F1}\,and\,\ref{F2}, and fits to all the observed spectra can be
found in the online Appendix\,B (Figs\,B1--B13).  The formal fitting errors
range from hundredths up to nearly $\sim$0.2\,dex with maximum values in the
case of titanium and nickel.  As was shown in Papers\,I--III, the
uncertainties in the abundances come mainly from uncertainties in stellar
parameters.  To examine how changes in stellar parameters affect the derived
abundances we performed additional calculations with atmospheric parameters
varied by typical values of their uncertainty: $\Delta T_{\rm{eff}} =
\pm$100\,K, $\Delta \log{g} = \pm$0.5 and $\Delta \xi_{\rm{t}} = \pm$0.25. 
The resulting values are shown in Table\,\ref{T5}.  The final estimated
uncertainty for each element is the quadrature sum of uncertainties of each
model parameter ($[(\Delta T_{\rm{eff}})^2 + (\Delta \log{g})^2 + (\Delta
\xi_{\rm{t}})^2]^{0.5}$) and it ranges from $\sim\pm$0.1\,dex up to
$\pm$0.25\,dex for the case of titanium.

\section{Concluding discussion}

The derived  metallicities cover a wide range from significantly subsolar to
slightly supersolar ([Fe/H] $= -1.0$ to $+0.6$ dex).  The histogram in
Fig.\,\ref{F5} shows the number of objects as a function of solar iron
abundance.  The maximum of the distribution is slightly subsolar, around
[Fe/H$]\sim-0.2$\,dex.  This reinforces our previous result (Paper\,III)
that symbiotic giants in general have subsolar metallicities with enhanced
mass-loss being responsible for high SySt activity.

\begin{figure}
   \includegraphics[width=\hsize]{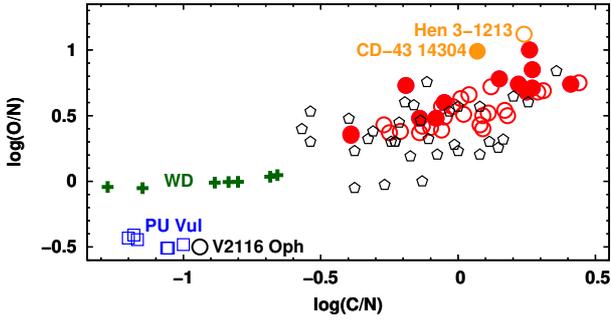}
      \caption{O/N versus C/N from the photospheric abundances.  Current
sample (filled circles) and sample from Paper\,III (open circles) compared
with values from nebular lines (pentagons) and for PU\,Vul during outburst
(squares).  Crosses represent theoretical predictions for nova ejecta from
CO WD with 0.65\,M\sun.}
         \label{F6}
   \end{figure}

All the symbiotic giants have enhanced $^{14}$N abundances, a signature that
these stars have experienced first dredge-up.  In Fig.\,\ref{F6}, the
photospheric O/N and C/N ratios from this paper and Paper\,III are compared
with values derived from nebular lines (\citealt{Nus1988},
\citealt{ScSc1990}, \citealt{Per1995}, \citealt{Sch2006}).  The 'nebular'
data are more spread in this plane probably as the result of larger
uncertainties.  The 'nebular' sample as a whole is moved with respect to
the 'photospheric' sample towards lower O/N and C/N ratios.  Extreme shifts
are seen by the values for symbiotic nebulae in outburst, represented here
by PU\,Vul during outburst \citep{VoNu1992}, and theoretical values for
ejecta from a 0.65 M\sun\, CO WD \citep{KoPr1997}.  Our current sample shows
somewhat lower nitrogen abundances with respect to that of the previous
sample (Paper\,III).  The difference is approximately $\sim$0.2--0.3\,dex. 
This possibly arises from the use in this paper of only spectra at
$\sim1.56$\,{\rm \micron} while Paper\,III had additional spectra at
$\sim2.23$\,{\rm \micron}.

The two alpha elements measured here, titanium and oxygen, are important in
understanding the formation and chemical evolution of Galactic populations. 
One route to studying the evolution of this pair of elements is to analyse
their abundances relative to iron.  Alpha elements and iron are produced in
different ways.  Alpha elements are created mostly by SNe\,II explosions of
massive stars and consequently have a relatively short time-scale.  Iron is
created by SNe\,Ia that span much longer time-scales.  The contamination of
the interstellar medium by these two constituents leads to different
chemical profiles for various populations.  Separate sequences can be
observed for distinct populations in diagrams constructed using the relative
abundances of alpha elements and iron.

\begin{figure}
   \includegraphics[width=\hsize]{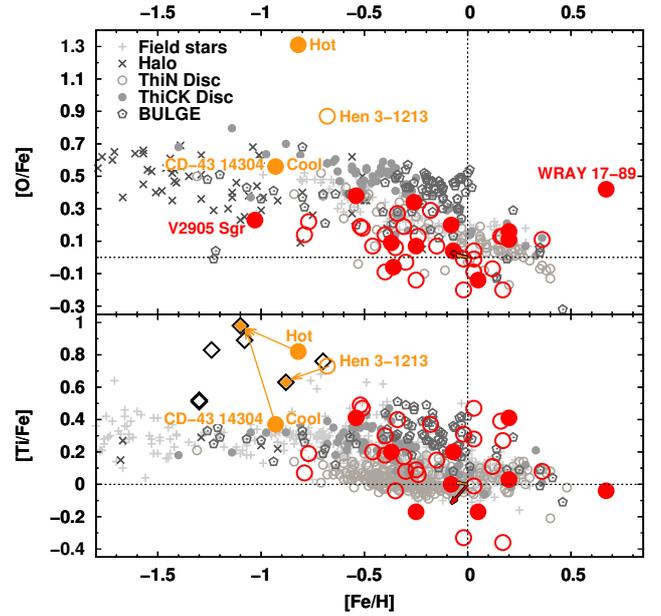}
      \caption{{\sl Top:} oxygen relative to iron for various stellar
populations with the positions of our targets denoted with large coloured
circles ('filled' for current sample and 'open' from Paper\,III).  For
CD-43$^{\circ}$14304 two cases are distinguished corresponding to solutions
with different stellar parameters: 'Cool' ($T_{\rm{eff}}$\,=\,3900\,K) and
'Hot' ($T_{\rm{eff}}$\,=\,4250\,K).  Four populations are distinguished:
thin and thick discs, halo and bulge.  {\sl Bottom:} titanium relative to
iron.  Black open diamonds are giants in yellow symbiotic systems from
\citet[1997]{Smi1996}, \citet{Per1998} and \citet{PeRo2009}.  Coloured
arrows starting from the 0,0 point in the coordinates show shifts to our
results applicable for the case with microtubulence velocity increased by
+0.25 km\,s$^{-1}$ (red and yellow for the M-type and the yellow giants,
respectively) according to the values given in the Table\,\ref{T5}.  See
online edition for colour version.}
         \label{F7}
   \end{figure}

Fig.\,\ref{F7} presents relative abundances [O$/$Fe] and [Ti$/$Fe] as a
function of metallicity ([Fe$/$H]) from our current (Table\,A1 in the online
Appendix\,A) and previous samples of symbiotic giants.  We also show values
for stars of various stellar populations (halo, thin and thick disc and
bulge) taken from a number of studies (see Paper\,III).  The position of our
objects in the [O$/$Fe] versus [Fe$/$H] diagram (Fig.\,\ref{F7} -- {\sl
top}) indicates that most SySt are the members of Galactic disc.  In the
cases when [Fe$/$H] is close to $-0.8$\,dex or smaller the extended
thick-disc/halo population is suggested.

In the [Ti$/$Fe] versus [Fe$/$H] plane (Fig.\,\ref{F7} -- {\sl bottom}) our
M-type giants are grouped mostly at higher [Ti$/$Fe] in the region
containing thick-disc and bulge stars.  Each of our samples contains one
yellow SySt, Hen\,3-1213 from Paper\,III and CD-43$^{\circ}$14304 from the
current paper.  These two stars show a small [Fe$/$H] and a large
enhancement of [Ti$/$Fe] suggesting membership in the halo.  Titanium and
iron abundances have been published by other authors for both stars as well
as for five other yellow symbiotics \citep[][AG\,Dra]{Smi1996},
\citep[][BD-21$^{\circ}$3873]{Smi1997}, \citep[][Hen\,2-467]{Per1998},
\citep[][CD-43$^{\circ}$14304, Hen\,3-863, StH$\alpha$\,176 and
Hen\,3-1213]{PeRo2009}.  In all these cases, similar enhancement of [Ti/Fe]
was found.  \citet{PeRo2009} also concluded that the overall abundance
pattern of yellow SySt follows the halo abundances.  The abundances obtained
for Hen\,3-1213 in our Paper\,III remain in good agreement with those of
\citet{PeRo2009}.  For CD-43$^{\circ}$14304, \citet{PeRo2009} obtained metal
abundances based on optical spectra as follows:
$\log{\epsilon(\mbox{Fe\,{\sc i}})}$ = 6.37 $\pm$ 0.19,
$\log{\epsilon(\mbox{Ni\,{\sc i}})}$ = 5.05 $\pm$ 0.26 and
$\log{\epsilon(\mbox{Ti\,{\sc i}})}$ = 4.81.

In our study, we adopted $T_{\rm{eff}}$ = 3900\,K for CD-43$^{\circ}$14304
corresponding to spectral type K7 according to the \citet{MuSm1999}
classification and in agreement with the temperature derived from infrared
colours (see Table\,\ref{T2}).  Taking into account fitting errors and
uncertainties in atmospheric parameters the obtained iron abundance
(6.54$\pm$0.10) remains in good agreement with that derived by
\citet{PeRo2009}.  However, \citet{PeRo2009} adopted a 400 K warmer
$T_{\rm{eff}}$, 4300\,K.  While the change in temperature does not
significantly affect the resulted metallicity the derived titanium abundance
(4.37$\pm$0.20) is much lower, shifting the position of this star in
[Ti$/$Fe] versus [Fe$/$H] diagram into the region of stars characterized by
typical compositions (Fig.\,\ref{F7} -- {\sl bottom}).  As a check, we
modelled CD-43$^{\circ}$14304 with a hotter $T_{\rm{eff}}$, 4250\,K, giving
atmospheric parameters similar to those obtained by \citet{PeRo2009}:
$T_{\rm{eff}}$ = 4300\,K and $\log{g}$=+1.6.  In this case, the iron
abundance (6.65$\pm$0.09) also agrees and the titanium (4.93$\pm$0.20) is an
almost perfect match to \citet{PeRo2009}.  However, the resulting unusually
high abundance of oxygen (9.18$\pm$0.24) would lead to very unrealistic
ratio of [O$/$Fe] $\sim +1.3$ (Fig.\,\ref{F7} -- {\sl top}).

\citet{PeRo2009} in their study determined the effective temperature using
the classical technique of comparing neutral and ionized iron line
abundances in optical spectra.  The spectrum used for CD-43$^{\circ}$14304
was obtained on 2007 August 26 (JD$\sim$2454339).  According to ephemeris
for eccentric orbit by \citet{Schm1998}, it corresponds to orbital phase
$\sim$0.09.  This phase is close to inferior spectroscopic conjunction with
giant in the front.  However, phase 0.09 is also soon after periastron
passage when strengthened interactions between the binary members cause an
increase in the nebular continuum that would strongly affect the depth of
absorption lines.  Indeed \citet{Gro2013} observed periodic brightening in
$V$ light curve in this system related to the periastron passage.

This example shows that considerable caution should be used when assigning
effective temperatures to giants in SySt for chemical composition
calculations.  Abundances of some elements strongly depend on the
temperature, as we have illustrated here for titanium and oxygen in hot,
yellow giants (see Table\,\ref{T5}).  Previously (Paper\,III) we reported
confirmation of high titanium abundances in yellow symbiotic systems and as
well as a trend of increasing titanium with decreasing metallicity for both
types of SySt, those with yellow and red M-type giants.  In light of the
example of CD-43$^{\circ}$14304 this deserves additional study with
temperatures reliably determined from infrared spectra.  In at least some
cases, the high titanium abundance could result from adopting an incorrectly
high effective temperature.

   \begin{figure}
   \includegraphics[width=\hsize]{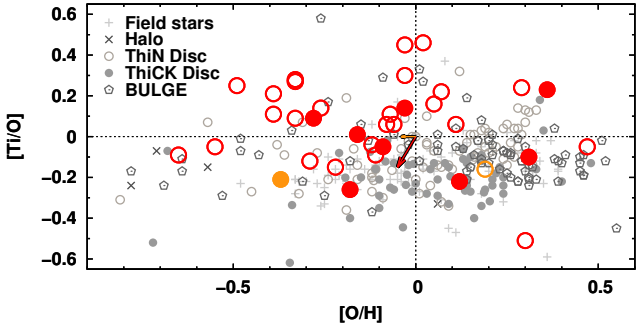}
      \caption{Titanium relative to oxygen for various stellar populations. 
Symbols analogous to those in Fig.\,\ref{F7}.}
         \label{F8}
   \end{figure}

In Fig.\,\ref{F8}, the relative abundance of titanium to oxygen is presented. 
In this case the, vast majority of red giants are located at high [Ti$/$O] in
a region mainly characteristic of the bulge population.  Such increased
titanium abundances are difficult to explain by theoretical models of
nucleosynthesis in stellar interiors.  A more reasonable explanation is that
the microturbulent velocities used in our calculations, fixed on value 2
km\,s$^{-1}$ (see {\sl Section}\,\ref{AR}), are too small.  In
Fig.\,\ref{F7}\,and\,\ref{F8}, the coloured arrows at the centre of
the coordinates show shifts to obtained values that would result from
the microturbulent velocity increased by +0.25 km\,s$^{-1}$, the estimated
uncertainty in this parameter (Table\,\ref{T5}).  The use of increased
microturbulence lowers the titanium overabundance observed in red symbiotic
giants.  We plan a further investigation into the microturbulence in the
near future using a selected sample of systems with well-known absolute
magnitudes.

\section*{Acknowledgements}
This study has been financed by the NCN postdoc programme FUGA (CG) via
grant DEC-2013/08/S/ST9/00581 and partly financed with NCN grant
DEC-2011/01/B/ST9/06145.  The high-resolution spectra were obtained with the
NOAO Phoenix spectrograph used at the Gemini Observatory.  The Gemini
Observatory is operated by the Association of Universities for Research in
Astronomy, Inc., under a cooperative agreement with the NSF on behalf of the
Gemini partnership: the National Science Foundation (United States), the
National Research Council (Canada), CONICYT (Chile), the Australian Research
Council (Australia), Minist\'{e}rio da Ci\^{e}ncia, Tecnologia e
Inova\c{c}\~{a}o (Brazil) and Ministerio de Ciencia, Tecnolog\'{i}a e
Innovaci\'{o}n Productiva (Argentina).  Observations of low-resolution
spectra were obtained with 1.9 m telescope at the South African Astronomical
Observatory.








\newpage
\clearpage
\newpage
\appendix

\section{Relative abundances}
\setcounter{page}{9}

\newpage

\begin{table}
  \caption{Selected relative abundances adopted for comparison with
Galactic stellar populations (Figures\,7\,and\,8).}
  \label{TA1}
  \begin{tabular}{@{}lccccc@{}}
  \hline
   Object            & $[$O$/$Fe$]$ & $[$Ti$/$Fe$]$ & $[$Ti$/$O$]$ & $[$Fe$/$H$]$ & $[$O$/$H$]$ \\
  \hline
WRAY\,15-1470        & $+0.34$      &  --           &   --         & $-0.26$      &  $+0.08$    \\
Hen\,3-1341          & $+0.38$      & $+0.41$       & $+0.03$      & $-0.54$      &  $-0.16$    \\
WRAY\,17-89          & $+0.42$      & $-0.04$       & $-0.46$      & $+0.67$      &  $+1.09$    \\
V2416\,Sgr           & $+0.16$      & $+0.41$       & $+0.25$      & $+0.20$      &  $+0.36$    \\
V615\,Sgr            & $+0.04$      & $+0.20$       & $+0.16$      & $-0.07$      &  $-0.03$    \\
AS\,281              & $-0.06$      &  --           &   --         & $-0.36$      &  $-0.42$    \\
V2756\,Sgr           & $+0.20$      & $+0.00$       & $-0.20$      & $-0.08$      &  $+0.12$    \\
V2905\,Sgr           & $+0.23$      &  --           &   --         & $-1.03$      &  $-0.80$    \\
AR\,Pav              & $+0.07$      & $-0.17$       & $-0.24$      & $-0.26$      &  $-0.19$    \\
V3804\,Sgr           & $-0.14$      & $-0.17$       & $-0.03$      & $+0.05$      &  $-0.09$    \\
V4018\,Sgr           & $+0.11$      & $+0.03$       & $-0.08$      & $+0.20$      &  $+0.31$    \\
V919\,Sgr            & $+0.09$      & $+0.20$       & $+0.11$      & $-0.37$      &  $-0.28$    \\
CD$-$43\degr14304    & $+0.56$      & $+0.37$       & $-0.19$      & $-0.93$      &  $-0.37$    \\
  \hline
\end{tabular}
\end{table}  

\clearpage

\section{Spectra of all, 13 symbiotic giants, compared with synthetic fits}

\newpage

   \begin{figure}
   \includegraphics[width=\hsize]{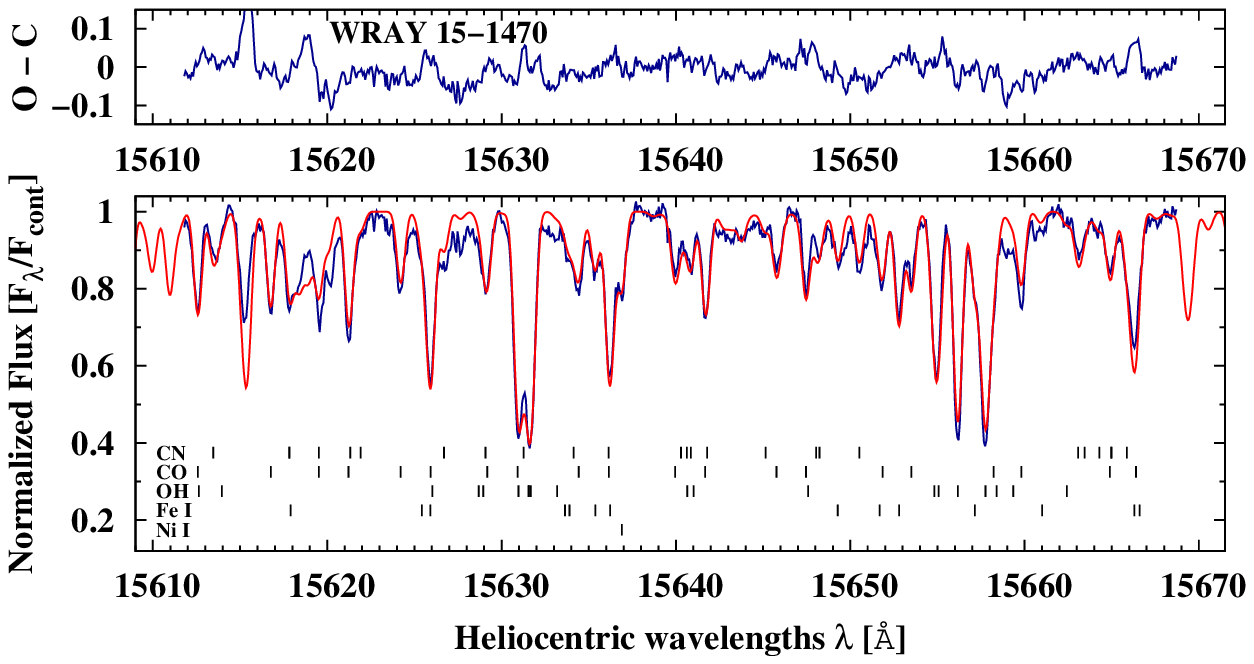}
      \caption{The spectrum of WRAY\,15-1470 observed in 2010 May (blue line)
              and a synthetic spectrum (red line) calculated using the final
              abundances (Table\,4).}
         \label{FB1}
   \end{figure}

   \begin{figure}
   \includegraphics[width=\hsize]{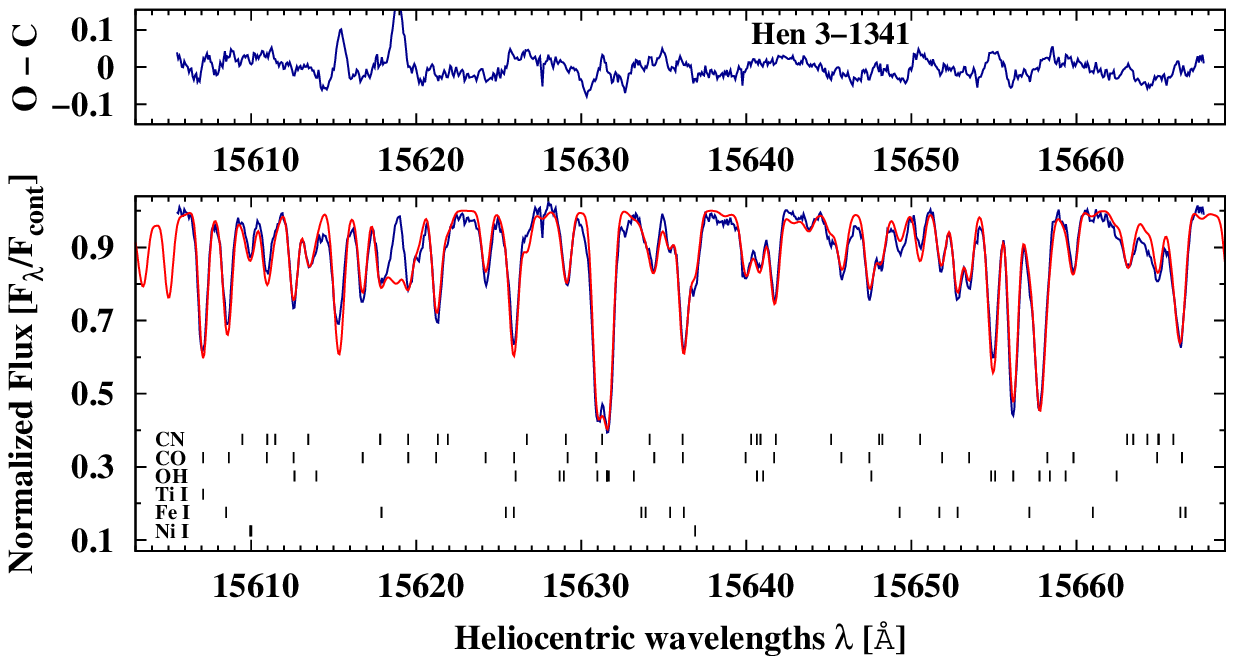}
      \caption{The spectrum of Hen\,3-1341 observed in 2010 May (blue line) and
              a synthetic spectrum (red line) calculated using the final
              abundances (Table\,4).}
         \label{FB2}
   \end{figure}

   \begin{figure}
   \includegraphics[width=\hsize]{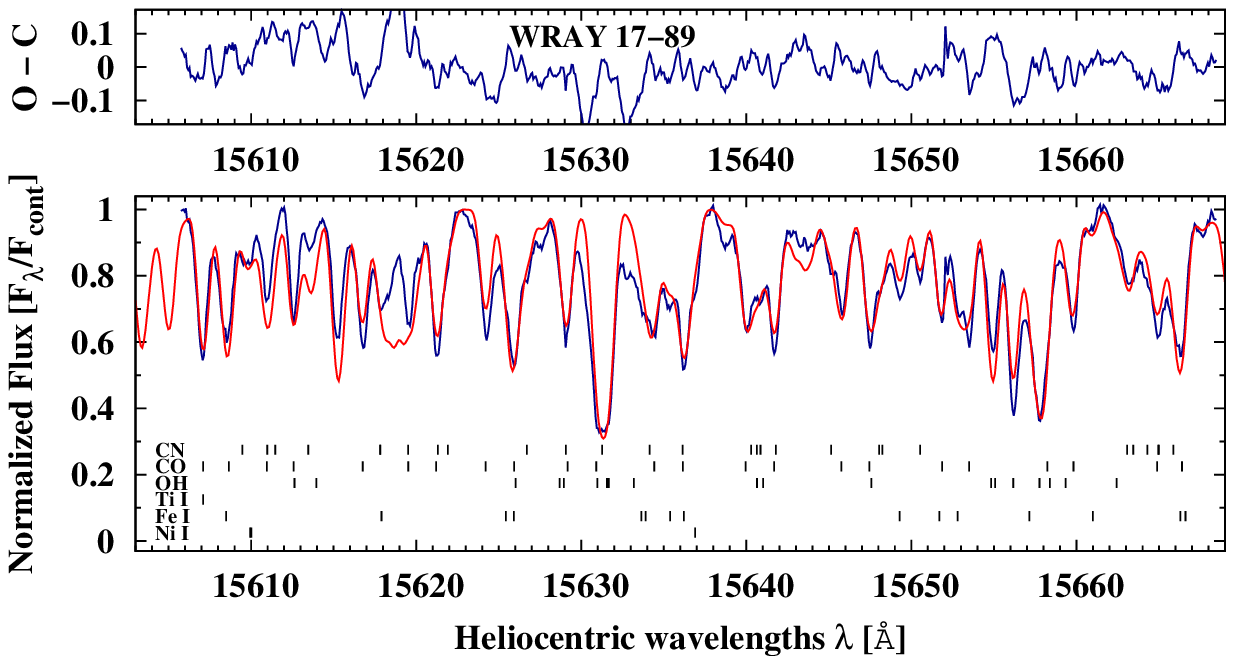}
      \caption{The spectrum of WRAY\,17-89 observed in 2010 May (blue line) and
              a synthetic spectrum (red line) calculated using the final
              abundances (Table\,4).}
         \label{FB3}
   \end{figure}

   \begin{figure}
   \includegraphics[width=\hsize]{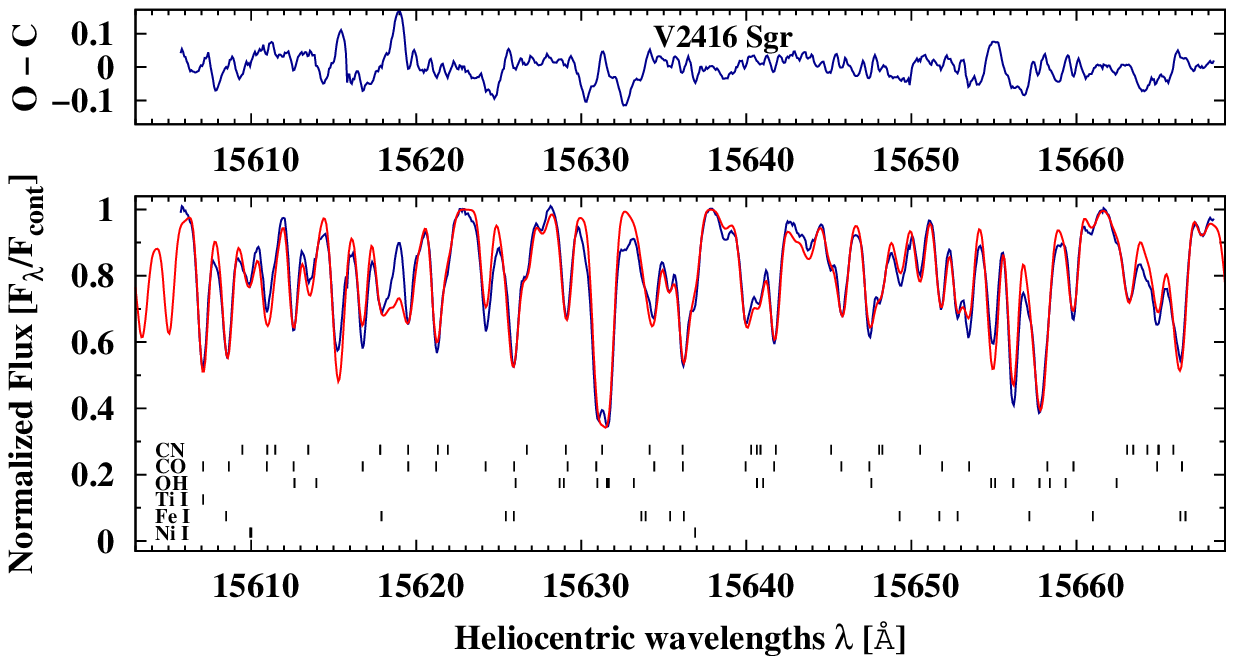}
      \caption{The spectrum of V2416\,Sgr observed in 2010 June (blue line) and
              a synthetic spectrum (red line) calculated using the final
              abundances (Table\,4).}
         \label{FB4}
   \end{figure}

   \begin{figure}
   \includegraphics[width=\hsize]{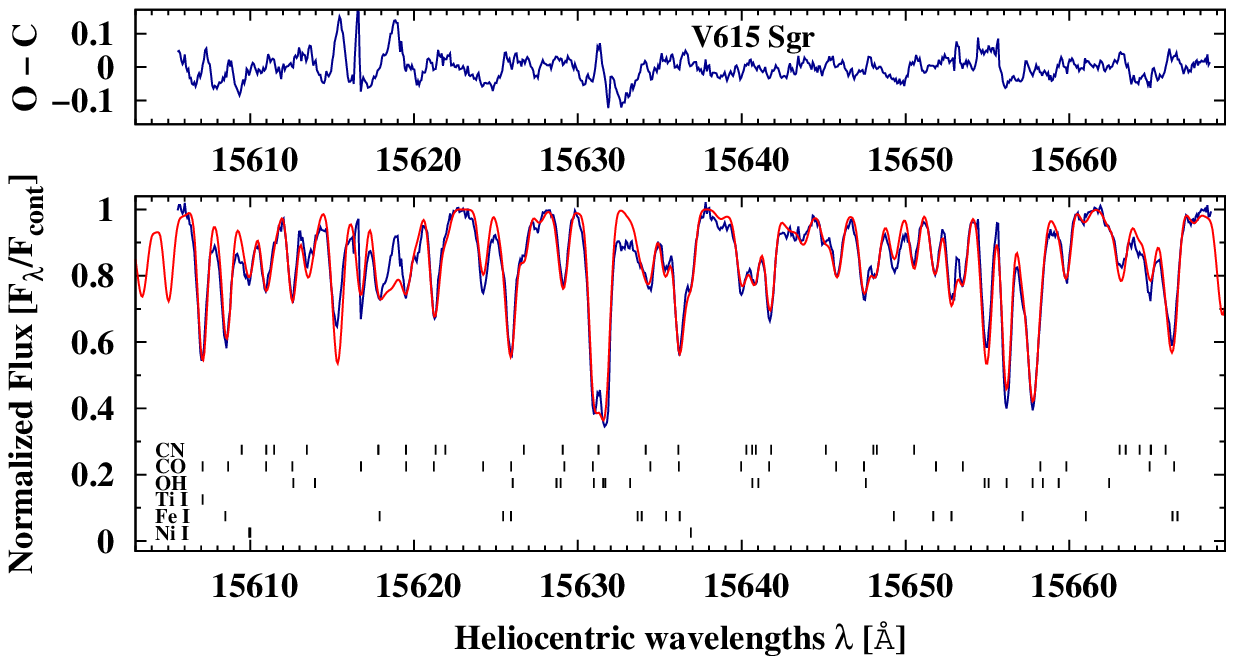}
      \caption{The spectrum of V615\,Sgr observed in 2010 May (blue line) and a
              synthetic spectrum (red line) calculated using the final
              abundances (Table\,4).}
         \label{FB5}
   \end{figure}

   \begin{figure}
   \includegraphics[width=\hsize]{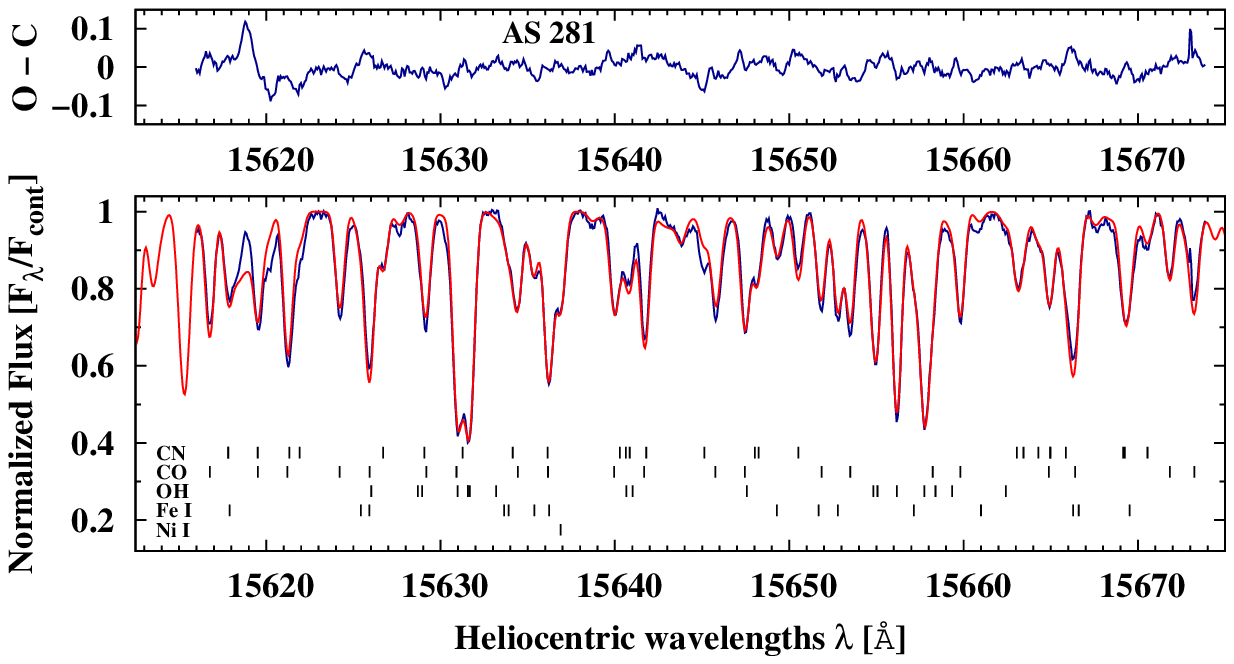}
      \caption{The spectrum of AS\,281 observed in 2010 June (blue line) and a
              synthetic spectrum (red line) calculated using the final
              abundances (Table\,4).}
         \label{FB6}
   \end{figure}

   \begin{figure}
   \includegraphics[width=\hsize]{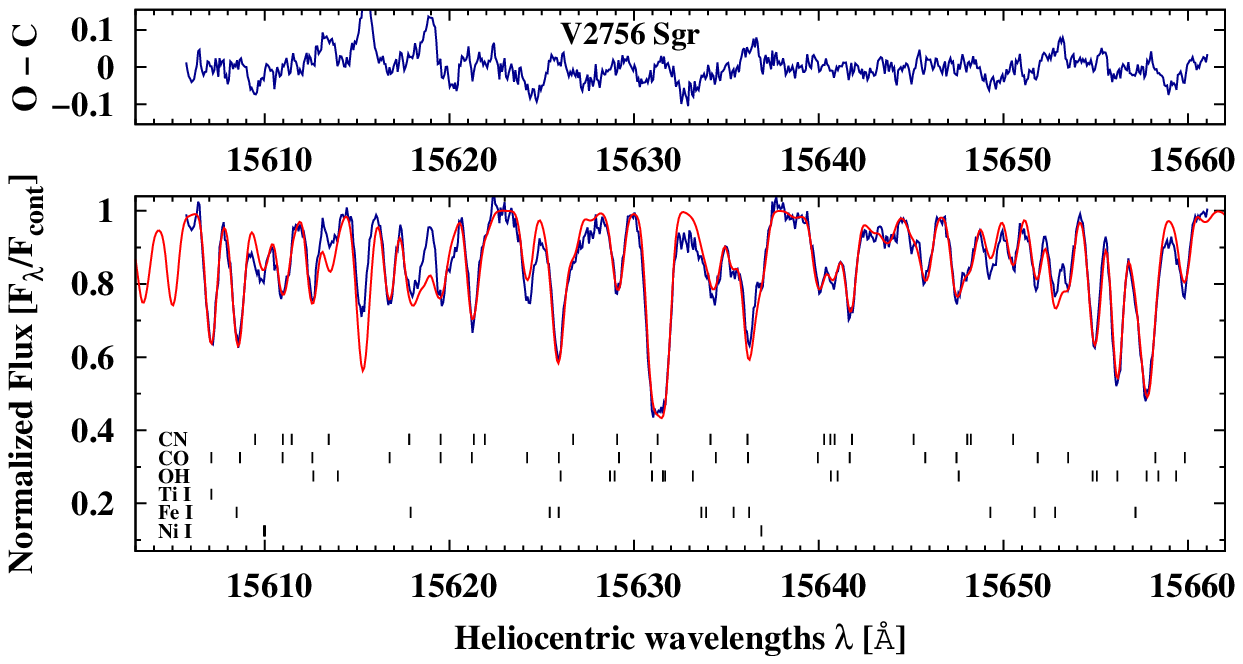}
      \caption{The spectrum of V2756\,Sgr observed in 2010 June (blue line) and
              a synthetic spectrum (red line) calculated using the final
              abundances (Table\,4).}
         \label{FB7}
   \end{figure}

   \begin{figure}
   \includegraphics[width=\hsize]{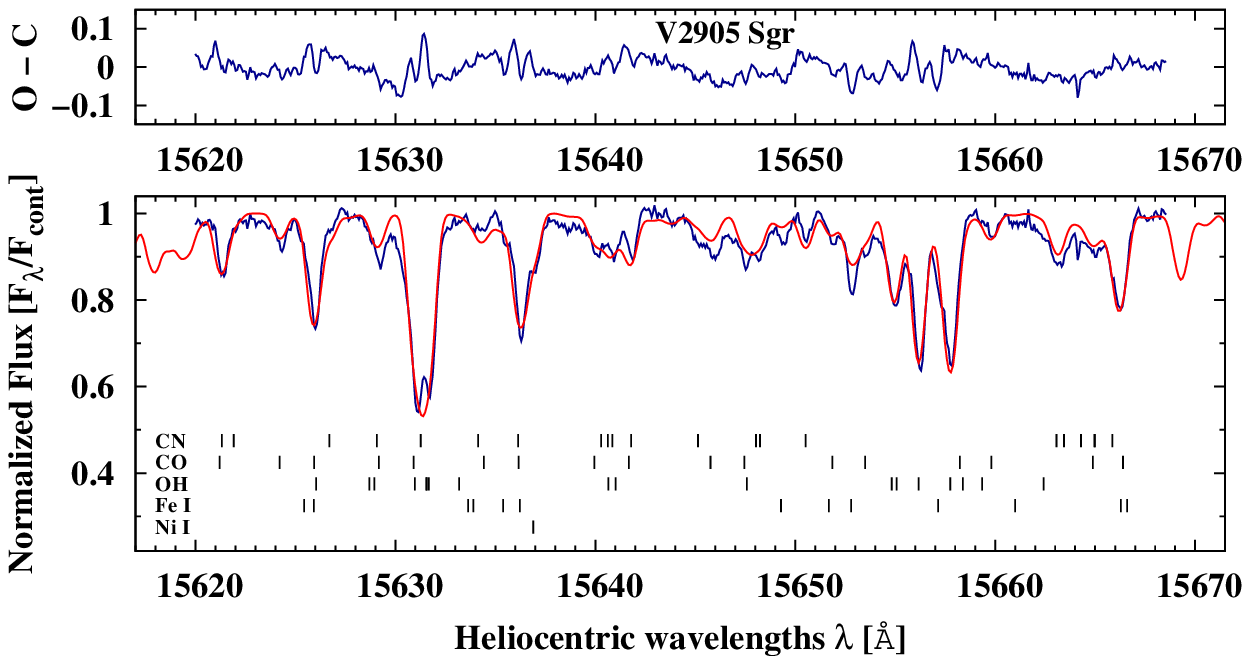}
      \caption{The spectrum of V2905\,Sgr observed in 2010 June (blue line) and
              a synthetic spectrum (red line) calculated using the final
              abundances (Table\,4).}
         \label{FB8}
   \end{figure}

   \begin{figure}
   \includegraphics[width=\hsize]{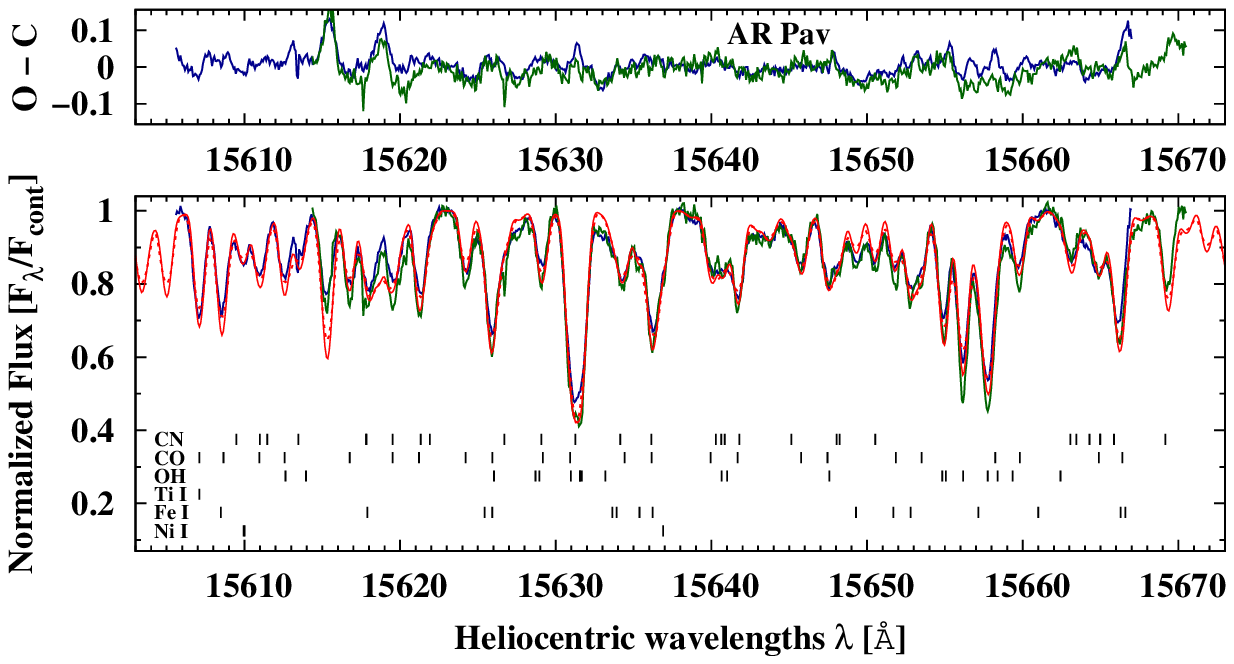}
      \caption{Spectra of AR\,Pav observed in 2009 June (blue line), 2010 May
              (green line), and a synthetic spectra (red continuous and
              dashed lines) calculated using the final abundances
              (Table\,4).}
         \label{FB9}
   \end{figure}

   \begin{figure}
   \includegraphics[width=\hsize]{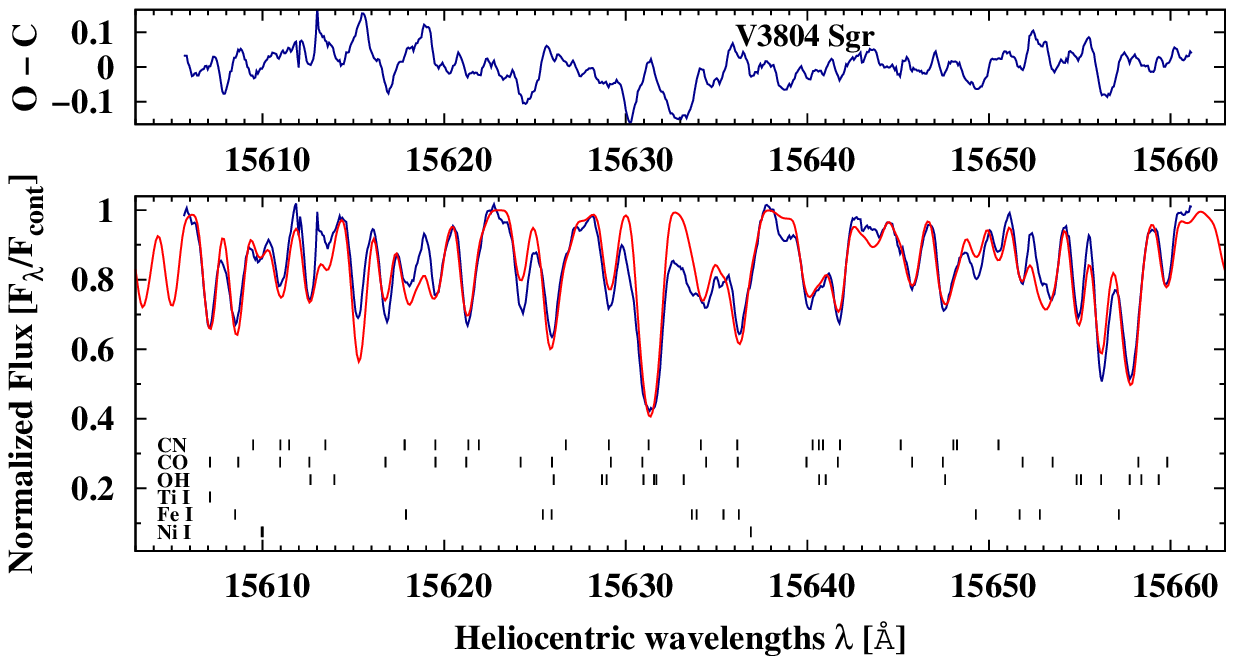}
      \caption{The spectrum of V3804\,Sgr observed in 2009 August (blue line)
              and a synthetic spectrum (red line) calculated using the final
              abundances (Table\,4).}
         \label{FB10}
   \end{figure}

   \begin{figure}
   \includegraphics[width=\hsize]{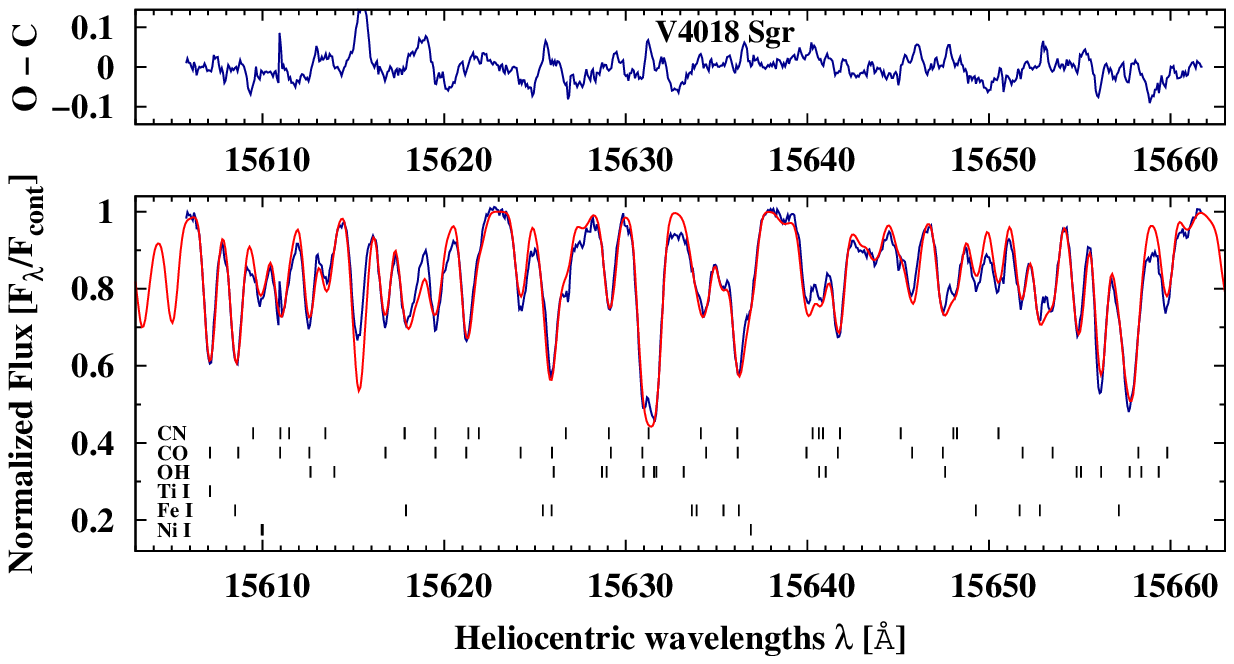}
      \caption{The spectrum of V4018\,Sgr observed in 2010 June (blue line) and
              a synthetic spectrum (red line) calculated using the final
              abundances (Table\,4).}
         \label{FB11}
   \end{figure}

   \begin{figure}
   \includegraphics[width=\hsize]{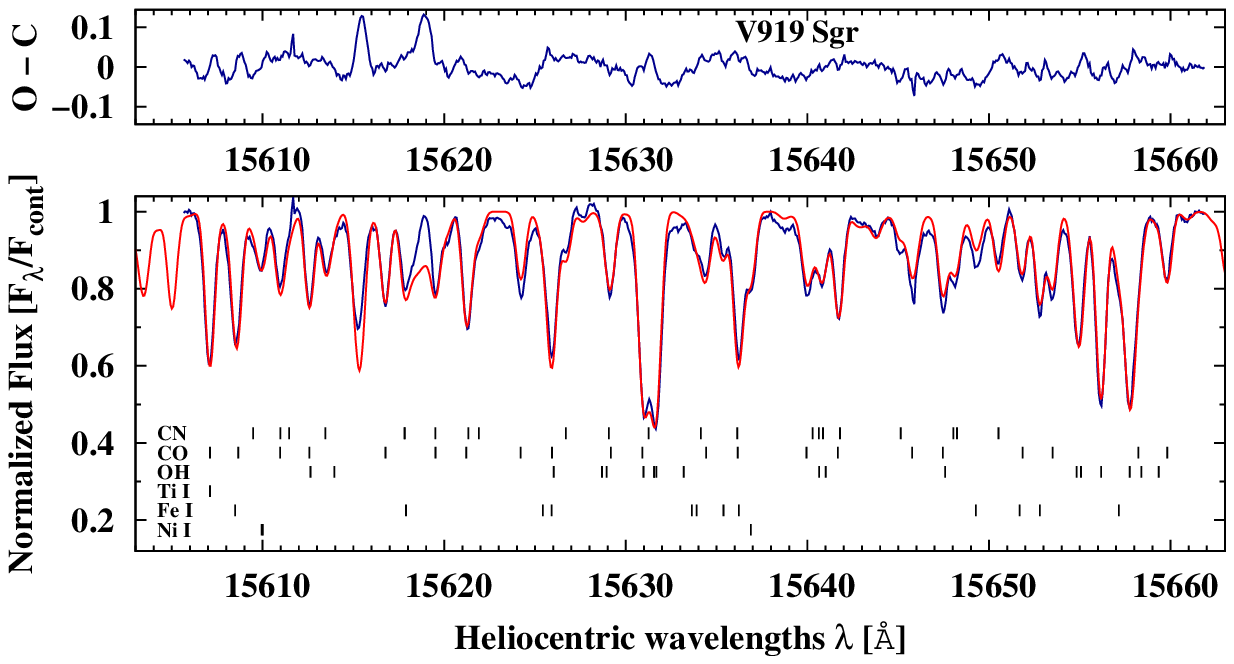}
      \caption{The spectrum of V919\,Sgr observed in 2009 August (blue line)
              and a synthetic spectrum (red line) calculated using the final
              abundances (Table\,4).}
         \label{FB12}
   \end{figure}

   \begin{figure}
   \includegraphics[width=\hsize]{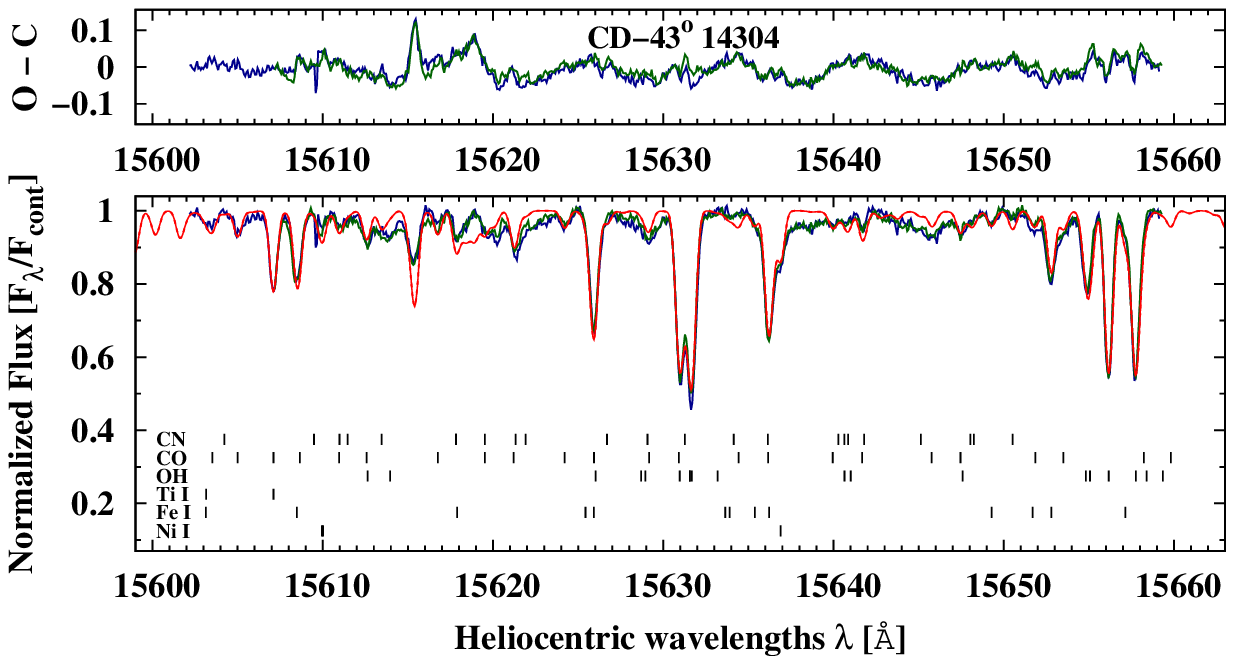}
      \caption{Spectra of CD$-$43\degr14304 observed in 2009 June (blue line),
              2010 May (green line), and a synthetic spectra (red continuous
              and dashed lines) calculated using the final abundances
              (Table\,4).}
         \label{FB13}
   \end{figure}


\bsp	
\label{lastpage}
\end{document}